\documentclass[referee,a4paper,12pt,traditabstract]{swsc} 

\usepackage{graphicx}
\usepackage{txfonts}
\usepackage{subfigure}
\usepackage{epstopdf}
\usepackage[authoryear,round]{natbib}
\usepackage[backref]{hyperref}
\usepackage{url}

\hypersetup{colorlinks=true,citecolor=cyan,urlcolor=cyan,linkcolor=blue}

\usepackage{gensymb}
\usepackage{siunitx}
\usepackage[table,xcdraw]{xcolor}
\usepackage{multirow}
\usepackage{color}
\usepackage{upgreek}
\newcommand{\Rs}{\(R_\odot\)}
\newcommand{\mymu}{$\upmu$}
\newcommand{\mylambda}{$\uplambda$}

\begin{document}

   \title{SunCET: The Sun Coronal Ejection Tracker Concept}
   
   \titlerunning{SunCET: A compact EUV instrument to fill a critical observational gap}

   \authorrunning{Mason}

   \author{James Paul Mason
          \inst{1}
          \and
          Phillip C. Chamberlin
          \inst{1}
          \and
          Daniel Seaton
          \inst{2} 
          \and
          Joan Burkepile
          \inst{3}  
          \and
          Robin Colaninno
          \inst{4} 
          \and
          Karin Dissauer
          \inst{5}  
          \and
          Francis G. Eparvier
          \inst{1}
          \and
          Yuhong Fan
          \inst{3}
          \and
          Sarah Gibson
          \inst{3}
          \and
          Andrew R. Jones
          \inst{1}
          \and
          Christina Kay
          \inst{6} 
          \and
          Michael Kirk
          \inst{6}
          \and
          Richard Kohnert
          \inst{1}
          \and
          W. Dean Pesnell 
          \inst{6}
          \and
          Barbara J. Thompson
          \inst{6}
          \and
          Astrid M. Veronig
          \inst{7}  
          \and
          Matthew J West
          \inst{8} 
          \and
          David Windt
          \inst{9}  
          \and
          Thomas N. Woods
          \inst{1}
          }

   \institute{Laboratory for Atmospheric and Space Physics, University of Colorado at Boulder,
              3665 Discovery Drive, Boulder, CO, USA\\
              \email{\href{mailto:james.mason@lasp.colorado.edu}{james.mason@lasp.colorado.edu}}
        \and
             NOAA/National Centers for Environmental Information, 
             325 Broadway, Boulder, CO, USA\\
        \and
             High Altitude Observatory, National Center for Atmospheric Research, 
             P.O. Box 3000, Boulder, CO, USA\\ 
        \and
             Naval Research Laboratory, 
             Washington, DC, USA\\
        \and
             Colorado Research Associates Division, NorthWest Research Associates, 
             3380 Mitchell Lane, Boulder, CO, USA\\
        \and
             NASA Goddard Space Flight Center, 
             8800 Greenbelt Road, Greenbelt, MD, USA\\
        \and
             Institute of Physics \& Kanzelh\"ohe Observatory for Solar and Environmental Research, University of Graz, 
             A-8010 Graz, Austria\\
        \and
             Royal Observatory of Belgium, 
             Avenue Circulaire 3, 1180 Uccle, Belgium\\
        \and
             Reflective X-ray Optics LLC, 
             New York, NY, USA\\
             }
 
  \abstract
   {The Sun Coronal Ejection Tracker (SunCET) is an extreme ultraviolet imager and spectrograph instrument concept for tracking coronal mass ejections through the region where they experience the majority of their acceleration: the difficult-to-observe middle corona. It contains a wide field of view (0--4~\Rs) imager and a 1~\AA\ spectral-resolution-irradiance spectrograph spanning 170--340~\AA. It leverages new detector technology to read out different areas of the detector with different integration times, resulting in what we call ``simultaneous high dynamic range", as opposed to the traditional high dynamic range camera technique of subsequent full-frame images that are then combined in post-processing. This allows us to image the bright solar disk with short integration time, the middle corona with a long integration time, and the spectra with their own, independent integration time. Thus, SunCET does not require the use of an opaque or filtered occulter. SunCET is also compact --- $\sim$15 $\times$ 15 $\times$ 10~cm in volume --- making it an ideal instrument for a CubeSat or a small, complementary addition to a larger mission. Indeed, SunCET is presently in a NASA-funded, competitive Phase A as a CubeSat and has also been proposed to NASA as an instrument onboard a 184 kg Mission of Opportunity.}


   \maketitle

\section{Introduction and Science Drivers}
\label{sec:intro}

The primary science question that the Sun Coronal Ejection Tracker (SunCET) instrument concept is designed to address is: \textit{What are the dominant physical mechanisms for coronal mass ejection acceleration as a function of altitude and time?}

\begin{figure}
\centering
\includegraphics[width=0.5\columnwidth]{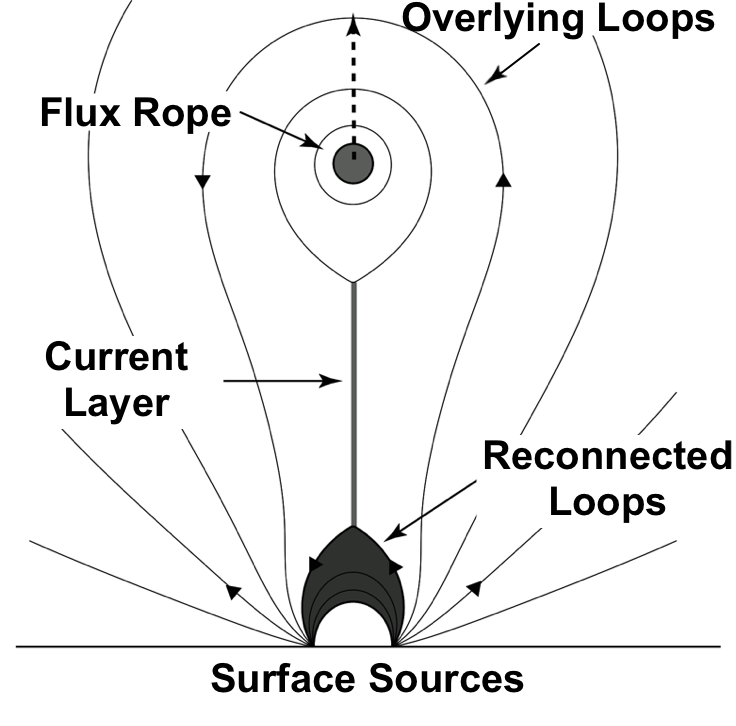}
\caption{\small Standard cartoon CME model. The flux rope extends through the page. Overlying fields resist the flux rope's elevation and expansion. Magnetic reconnection releases the energy stored in the field to accelerate the flux rope, producing a CME. Adapted from \citet{Forbes2018}.} 
\label{fig:cartoon_cme}
\end{figure}

\begin{figure}
\centering
\includegraphics[width=0.48\textwidth]{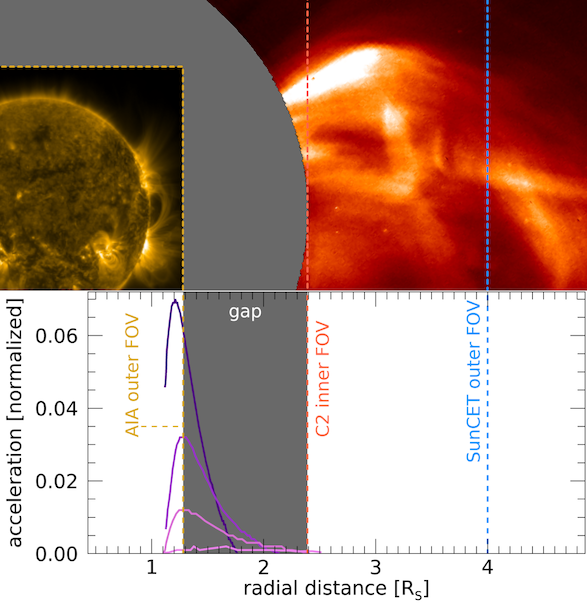}
\caption{Top: Composite of SDO/AIA 171 \AA\ image and SOHO/LASCO/C2 white-light coronagraph image. The longstanding observational gap is shown in dark grey. Bottom: Modeled acceleration profiles of torus instability CMEs, adapted from \citet{Torok2007} Fig. 3. The different curves result from different background magnetic field decay index profile assumptions, with each higher acceleration peak corresponding to a larger decay index profile. Most of the acceleration occurs in the observational gap that SunCET fills.}
\label{fig:one_liner}
\end{figure}

In the standard model configuration of a coronal mass ejection (CME; Figure \ref{fig:cartoon_cme}), a CME must overcome the constraint of overlying field in order to escape. Perhaps the simplest model of this defines a 1D, horizontal background magnetic field that declines in strength with height, characterized by the ``decay index" \citep{Bateman1978, Kliem2006}. If the background field decays too rapidly, the so-called torus instability of the embedded flux rope occurs, meaning the flux rope erupts. The decay index has a direct impact on the CME kinematics. The acceleration curves in the bottom of Figure \ref{fig:one_liner}, derived from magnetohydrodynamic (MHD) simulations by \citet{Torok2007}, correspond to decay index profiles, with each increase in acceleration corresponding to an increase in in the decay index profile and the final CME speed. Thus, the acceleration profile of a CME acts as a natural probe of the surrounding magnetic field. There are many complications layered on top of this simple model in reality, described later in this introduction.

The bulk of the CME acceleration profile in all cases occurs either in the observational gap or in the region where existing instruments are not optimized. This gap exists between extreme ultraviolet (EUV) imagers (widest outer field of view [FOV] of 1.7 \Rs) and coronagraphs (typical inner FOV of 2.5 \Rs\ but effectively higher due to diffraction-degraded spatial resolution). Some instruments observe only part of the low-middle corona (Solar TErrestrial RElations Observatory [STEREO; \citealt{Kaiser2007}] / Coronagraph-1 [COR1; \citealt{Howard2008}], Geostationary Operational Environmental Satellite [GOES] / Solar Ultraviolet Imager [SUVI; \citealt{Martinez-Galarce2010}] , Project for On-Board Autonomy [PROBA2] / Sun Watcher with Active Pixels and Image Processing [SWAP; \citealt{Seaton2013}]). Some have low signal to noise in the middle corona (SUVI, SWAP). Some are ground-based with duty cycles $<20$\% (K-Cor). Some have limitations on cadence (COR1). SunCET, however, avoids all of these issues because it is specifically optimized for this study of CMEs. Directly observing the CME height-time profile through the whole low and middle corona allows the derivation of complete speed-time and acceleration-time profiles, and thus strong model constraints, requiring accurate modeling of the magnetic environment to obtain the observed profiles. Such constraints do not presently exist, but SunCET can provide them.

The torus instability is not the only mechanism involved in CME eruptions. Complicating factors are introduced by, e.g., the 3D structure of the erupting material and the surrounding magnetic field, by potential drainage of dense plasma, and by continued magnetic reconnection freeing more energy to drive the CME. The influence of these factors also evolve with altitude and time, as the CME dynamics play out. There have been at least 26 review papers on the topic over the last two decades (\citealt{Green2018}, and references therein) --- a testament to the sustained, intense interest in this topic.

\begin{figure}
\centering
\includegraphics[width=0.48\textwidth]{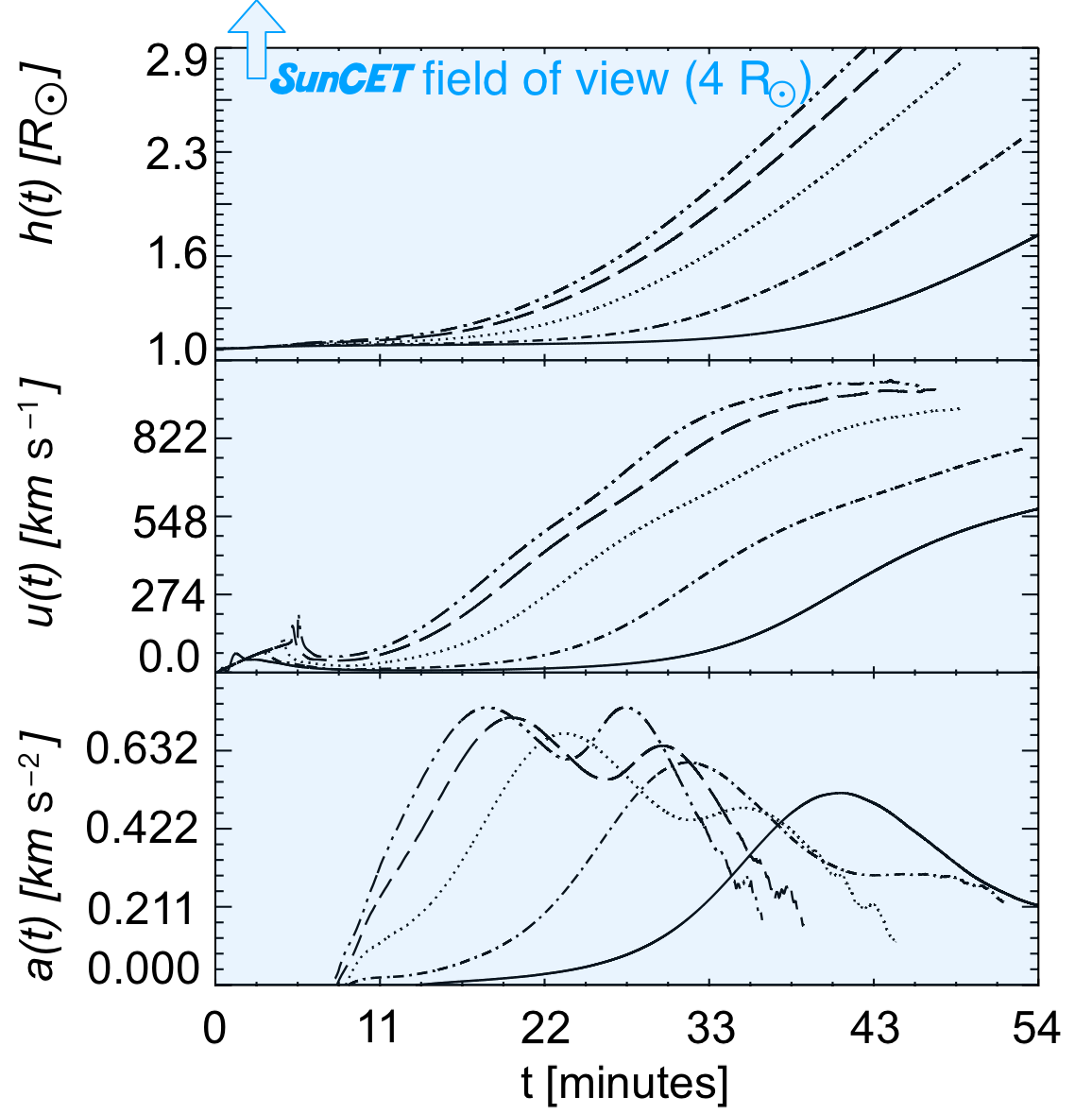}
\caption{Simulated CME kinematic profiles. Solid lines indicate the unperturbed torus instability. Dashed lines from right to left correspond to increasing durations (6 $\tau_A$ up to 10$\tau_A$) of an upward, linearly rising velocity perturbation, resulting in fundamentally different acceleration profiles. The SunCET FOV (0--4~\Rs; indicated in light blue) covers and extends beyond this simulation. Adapted from \citet{Schrijver2008} Fig. 7.}
\label{fig:torus}
\end{figure}

For example, a relatively modest complication to layer into the torus instability model is to add an upward velocity perturbation with finite duration. MHD simulations by \citet{Schrijver2008} showed that simply changing the duration of this perturbation results in fundamentally different acceleration profiles (Figure \ref{fig:torus}). With brief perturbations, the profile is single-peaked and occurs at later times. Increasing the duration of the perturbation does not simply result in an earlier peak, but in two peaks. Just as in Figure \ref{fig:one_liner}, the heights that these acceleration profiles differentiate themselves occurs across the Heliophysics System Observatory (HSO) measurement gap. SunCET observations can discriminate between single-peak versus double-peak CME acceleration profiles, which then determines the duration of a velocity perturbation in the torus instability model.

\begin{figure}
\centering
\includegraphics[width=0.68\textwidth]{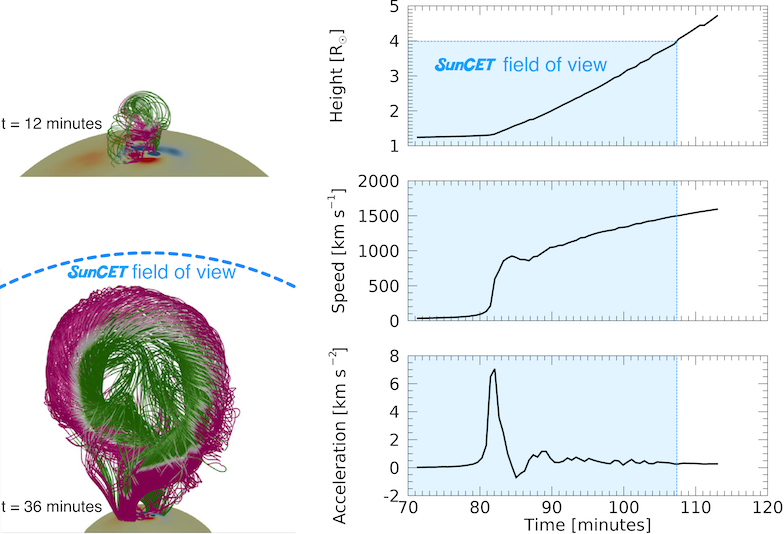}
\caption{MFE simulation containing the helical kink instability, resulting in impulsive CME acceleration. The SunCET FOV (0--4~\Rs; indicated in light blue) captures the impulse and small jerks. Adapted from \citet{Fan2016}.}
\label{fig:kink}
\end{figure}

Another CME initiation mechanism arises from the magnetic field topology of the flux rope. \citet{Hood1981} showed that if the total twist in a flux rope exceeds a critical threshold (448\degree), a ``helical kink" instability will occur, causing the flux rope to erupt. Such contortions lead to an impulsive acceleration and a large rotation of the flux rope (\citealt{Fan2016}, Figure \ref{fig:kink}). Note the substantial differences in the simulated acceleration profiles between Figures \ref{fig:one_liner}, \ref{fig:torus}, and \ref{fig:kink}; and that they all occur in the under observed region.

\begin{figure}
\centering
\includegraphics[width=0.7\textwidth]{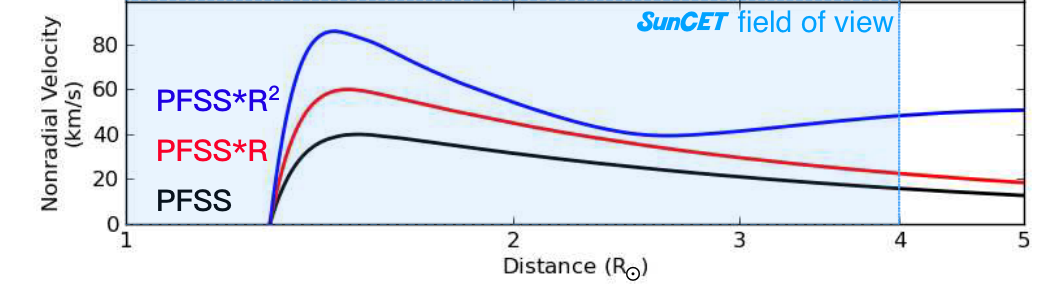}
\caption{ForeCAT simulations of a CME propagating through background magnetic fields (PFSS) of various strengths. R is radial distance. CMEs experience greater non-radial velocity in middle corona environments with stronger magnetic fields. The SunCET FOV (0--4~\Rs; indicated in light blue) captures the majority of CME deflection. Adapted from \citet{Kay2016}.}
\label{fig:deflection}
\end{figure}

The other aspect of acceleration is direction: CMEs can be deflected away from ``pure" radial propagation by as much as $\sim$30$\degree$, which is again determined primarily by \textbf{B}$_{ex}$ (Figure \ref{fig:deflection}). This force has a non-radial component because the field is not perfectly symmetric about the flux rope, causing a magnetic gradient on the CME's sides as the loops drape around the rising CME. The Forecasting a Coronal Mass Ejection's Altered Trajectory (ForeCAT) analytical model accounts for these and other forces on a CME to determine its non-radial velocity \citep{Kay2013, Kay2015a, Kay2016a, Kay2018}. Furthermore, \citet{Kay2015} modeled 200 CMEs in ForeCAT and found that deflection occurring in the middle corona accounts for nearly all of the deflection that occurs between initiation and 1 AU. The background magnetic field and radial CME speed are two free parameters in ForeCAT that are critical to get right; SunCET observations can strictly constrain them via forward modeling.

Additionally, coronal dimming often occurs as a result of CMEs. The faster a CME departs, the steeper the decline in coronal emission. The more mass the CME takes with it, the deeper the drop in coronal emission. A large number of studies have demonstrated this link with coronal imagers (e.g., \citealt{Aschwanden2009b, Aschwanden2009, Dissauer2018, Dissauer2019, Thompson2000}) and with spectral irradiance data \citep{Woods2011, Mason2014, Mason2016, Mason2019}. A major advantage of dimming measurements is that they are effective measures of CME kinematics even when they occur at disk center. Coronagraphs and imagers suffer from the problem of determining halo CME speed and/or mass. Dimming is an effective measure of CME kinematics both on and off-disk \citep{Dissauer2019,Chikunova2020}. Thus, instrument suites that can capture both the dimming and direct observations of limb CMEs are ideal for CME observation. This is precisely what SunCET does. 

SunCET will be the first mission that allows continuous measurements of CMEs during their initial acceleration phase using only a single instrument. This is advantageous compared to currently used instruments, where, e.g. EUV imagers in the low corona are combined with white-light coronagraphs higher up to track this phase. Artifacts can be introduced in the resulting CME kinematics using this combined data due to the tracking of different structures in the different instruments, since the observed emission is generated by different physical processes. SunCET is not dependent on other instruments to observe CME initiation and acceleration but does have a sufficiently wide field of view to overlap with coronagraphs for further expanded studies. The same challenges with different CME structures in EUV versus white light will be present, but SunCET's broader temperature response should mitigate this somewhat.

\section{Instrument Design}
\label{sec:technical}

SunCET is an instrument with a Ritchey-Chr\'etien, wide-field-of-view telescope (4 \Rs), an off-rowland-circle EUV spectrograph, and a novel, simultaneous-high-dynamic-range detector. This new detector technology allows us to image the bright solar disk and CMEs through the dim middle corona simultaneously. It also allows us to measure solar irradiance spectra on the unused portion of the same detector with an integration time independent of the telescope image. The entire design is compact, fitting in a $\sim$15 $\times$ 15 $\times$ 10~cm volume; or about 2.5 CubeSat Units. This makes it ideal as a CubeSat or as a compact instrument suite to include on larger spacecraft that requires few physical resources. 

SunCET observes in the EUV rather than white light because 1) CMEs have already been demonstrated to be visible in the EUV and 2) it allows for major simplifications in the technical design of the instrument. While white-light observations are independent of temperature since they rely on light Thomson scattered from free electrons, SunCET observations do have the caveat that their temperature dependence (emission from ions at particular temperatures) means that CMEs whose plasma is not at ambient coronal temperatures will not be visible. The dynamic range between on and off disk in the EUV is already large ($\sim$10$^5$ by 2 \Rs) but this is orders of magnitude larger in white light ($\sim$10$^8$), increasing the technical challenge. Moreover, the absolute brightness are vastly different; there are far more visible light photons. This presents a major challenge with scattered light: even small imperfections in optics would result in enough of the numerous disk photons to land on the part of the detector with the exceptionally faint middle corona, swamping out CME observations. This is further exacerbated by the fact that most surfaces scatter light more efficiently in visible light than in EUV light. Therefore, SunCET observes CMEs in the EUV.

\section{Imager Design}
\label{sec:imager_technical}
The SunCET imager was designed to provide high-dyamic range with moderate spatial resolution while providing a large field-of-view not heard of in historical on-disk EUV imagers out to 4 \Rs. This section describes the technical design details that were traded in order to close on the science question.

\subsection{Dynamic Range}
\label{sec:dynamic_range}

The SunCET imager requires a dynamic range of at least \num{7e4}, based on GOES-16/SUVI observations of CMEs and SunCET's design optimizations. The dimmest target of interest is a CME at the outer FOV, and the brightest is the coronal loops of an active region associated with a CME.

SUVI-observed radiances are used to estimate brightness in SunCET (see Section \ref{sec:snr}). At 3.5~\Rs, CMEs are \num{6.9e-4} W/m$^2$/sr. A few of the brightest pixels in active regions reach $\sim$70 W/m$^2$/sr, but are typically $\sim$4.8 W/m$^2$/sr in SunCET. Another factor of 10 is included to distinguish the loops from the background solar disk. Thus, we have a required dynamic range of (4.8 / \num{6.9e-4}) $\times$ 10 = \num{7e4}. We allow solar flares and a small number of the brightest pixels inside active regions to saturate because 1) they are not our target of interest, 2) our entrance filter mesh mitigates diffraction (Section \ref{sec:bandpass}), and 3) the blooming in our detector is modest: only a few percent ranging across a few pixels (verified during the 33.336 NASA sounding rocket flight and in the lab).

\textit{Projected performance}: CME brightness at the outer SunCET FOV of 4 \Rs\ is \num{2.1e-4} W/m$^2$/sr. That implies a dynamic range of \num{2.3e5}. From 0--1.05~\Rs, we run exposures of 0.025 seconds and from 1.05--4~\Rs\ the exposures will be 10 seconds --- a factor of 400$\times$ dynamic range. Our detector has a native dynamic range of $\sim$\num{5e3}. 2$\times$2 pixel binning provides an additional factor of 4. Combining these, we obtain SunCET's high dynamic range of \num{8e6}, well above the required range of \num{7e4}. For comparison, the SDO/AIA dynamic range is \num{1e4} \citep{Lemen2012}.

\subsection{Field of View}
\label{sec:fov}

Most CMEs accelerate through the low and middle corona \citep{Bein2011, DHuys2014}. We set our required minimum field of view (FOV) at 0.5 \Rs, corresponding to $\pm$30$\degree$ from disk-center. Lower than this and the events tend to be halo CMEs, which are difficult to obtain height-time profiles from. The outer FOV requirement is set to 3.5~\Rs. SunCET covers the gap between existing instruments and includes enough overlap to ensure a smooth transition in any complementary height-time profiles. SOHO/LASCO's inner FOV is 2.4~\Rs\ and its upcoming replacement, NOAA's GOES-U/CCOR and SWFO/CCOR, will have an inner FOV of 3~\Rs.  

The aforementioned traditional CME measurements, which are from white-light coronagraphs, use occulters that are mechanically restricted to be a limited distance away; therefore these observations have significantly degraded spatial resolution in their inner FOV that is much worse than their stated plate-scale resolution, sometimes upwards of 1 arc-min in the inner FOV. These effects are primarily due to vignetting (e.g. \citealt{Koutchmy1988, Aime2019}). This is not the case with SunCET as it does not require an occulter to observe the CMEs in the low- and middle-corona, so its spatial resolution is not diffraction limited and is superior even in the FOV region that overlaps with the coronagraphs. 

\textit{Projected performance}: The FOV of SunCET is 0--4~\Rs\ (5.6~\Rs\ in image corners).

\subsection{Temporal Resolution: Exposure and Cadence}
\label{sec:temporal_resolution}

SunCET is required to observe CMEs with speeds up to at least 1000 km/s, which accounts for 98\% of all CMEs \citep{Gopalswamy2009, Barlyaeva2018}. Given the cadence described below and the field of view, SunCET's projected performance is to observe CMEs with speeds up to 3900 km/s. The fastest CME in the CDAW catalog is $\sim$3400 km/s, meaning that SunCET will be able to track CMEs with any previously observed speed.

SunCET requires an exposure time $\leq$23 seconds in order to avoid motion blur of the CME. Combining the fastest required CME to observe (1000 km/s), our required spatial resolution of 30\arcsec/resolution-element, and the conversion of angular to spatial resolution at 1~AU ($\sim$750 km/arcsec), we obtain 750 $\times$ 30 / 1000 $\approx$ 23 seconds/resolution-element. 
\textit{Projected performance - exposure}: SunCET's exposure times are 0.025 seconds from 0--1.05~\Rs\ and 10 seconds beyond that.

SunCET requires a cadence $\leq$3.2 minutes. SunCET must be able to track a 1000 km/s CME from the solar limb through its FOV, a range of 2.5~\Rs, or \num{1.74e6} km. Therefore, the minimum time a CME would be in the FOV is 29 minutes. We require at least 9 height-time samples to distinguish acceleration profiles (Figure \ref{fig:torus}). Thus, our cadence must be less than 29 minutes / 9 samples = 3.2 minutes. 

\textit{Projected performance - cadence}: The SunCET mission is designed to downlink 1 minute cadence data. The designed FOV actually extends to 4 \Rs, meaning we will capture 38 height-time points for limb-CMEs traveling at a speed of 1000 km/s and more points for CMEs that start slightly on disk and/or with slower speeds. For example, the average CME speed is 490 km/s \citep{Webb2012} and if it crosses from 0.7--4~\Rs, we will obtain 78 height-time points.

\subsection{Bandpass: Coatings and Filters}
\label{sec:bandpass}

\begin{table}
\caption{Strong emission lines in the SunCET bandpass. Irradiance measured by SDO/EVE \citep{Woods2012}}
\label{tab:bandpass_lines}
\begin{tabular}{|c|c|c|c|}
\hline
\rowcolor[HTML]{1A73C9} 
{\color[HTML]{FFFFFF} Ion} & {\color[HTML]{FFFFFF} \mylambda\ [\AA]} & {\color[HTML]{FFFFFF} log$_{10}$(T [K])} & {\color[HTML]{FFFFFF} \begin{tabular}[c]{@{}c@{}}Quiet Sun \\ Irradiance \\\relax [\mymu W/m$^2$/\AA]\end{tabular}} \\ \hline
Fe IX & 171.1 & 5.9 & 67 \\ \hline
Fe X & 174.5 & 6.1 & 73 \\ \hline
Fe X & 177.2 & 6.1 & 48 \\ \hline
Fe XI & 180.4 & 6.2 & 77 \\ \hline
\begin{tabular}[c]{@{}c@{}}Fe XI\\ (doublet)\end{tabular} & 188.2 & 6.2 & 61 \\ \hline
Fe XII & 193.5 & 6.2 & 45 \\ \hline
Fe XII & 195.1 & 6.2 & 63 \\ \hline
\end{tabular}
\end{table}

CMEs have been routinely identified in narrowband EUV imagers sensitive to temperatures between $\sim$0.6--1.6~MK (e.g., GOES/SUVI). Therefore, SunCET is required to observe at least one of the emission lines identified in Table \ref{tab:bandpass_lines}.

\begin{figure}
\centering
\includegraphics[width=0.7\textwidth]{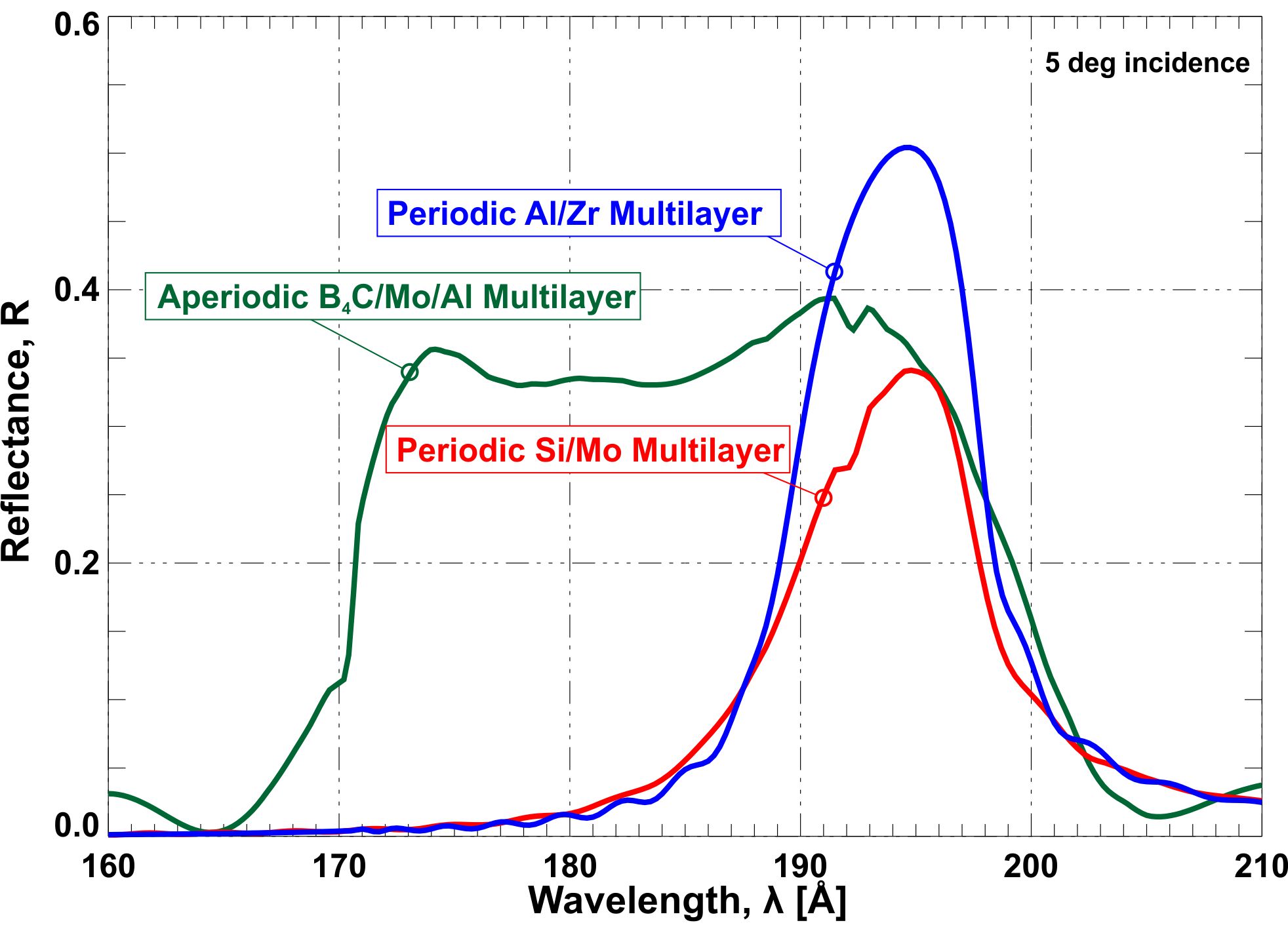}
\caption{Calculated reflectance near normal incidence (5\degree) of the broad-band, aperiodic B$_4$C/Mo/Al multilayers used for the SunCET telescope mirrors (green), and for reference, the narrow-band, periodic Si/Mo multilayer coatings used for the GOES/SUVI instrument (red), and the Al/Zr multilayer coatings used for the Hi-C rocket instrument (blue).}
\label{fig:coating}
\end{figure}

\textit{Projected performance}: SunCET's baseline bandpass is 170--200~\AA\ --- capturing all of the emission lines in Table \ref{tab:bandpass_lines}, which boosts the signal (Section \ref{sec:snr}). The telescope mirrors employ reflective multilayer coatings designed to provide broad spectral response spanning the instrument bandpass. These coatings follow an aperiodic design, and comprise 15 repetitions of alternating layers of B$_4$C, Mo, and Al, with individual layer thicknesses ranging from $\sim$5--100~\AA. The aperiodic coating design provides an average reflectance of $\sim$33\% from 170--200~\AA, as shown in Figure \ref{fig:coating}. For reference, periodic multilayer coatings operating in this portion of the EUV are generally used for narrow-band response: for example, the periodic Si/Mo coatings used for the 195~\AA\ channel of the GOES/SUVI instrument, also shown in Figure \ref{fig:coating}, achieve a peak reflectance of $\sim$34\% with a spectral bandpass of $\sim$9.5~\AA\ full-width-half-max (FWHM). Figure \ref{fig:coating} also shows the periodic Al/Zr coatings used for the Hi-C rocket instrument \citep{Kobayashi2014}, which achieve a peak reflectance of $\sim$50\% with a spectral bandpass of $\sim$8.5~\AA\ FWHM. The aperiodic B$_4$C/Mo/Al multilayer coatings are currently under development with funding from the NASA H-TIDeS program.

The C/Al/C entrance filter from Luxel Corporation prevents visible light from entering the chamber and has high heritage (24 of them in GOES/EXIS). It is supported by a 5 lines/inch mesh, which has heritage from the Hi-C sounding rocket flights and avoids the diffraction issues of the 70 lines/inch mesh used on SDO/AIA and TRACE \citep{Lemen2012, Lin2001}. A second C/Al/C filter directly in front of the detector eliminates visible light from possible pinholes in the primary filter or from stray light in the instrument.

\subsection{Spatial Resolution}
\label{sec:spatial}

\begin{figure}
\centering
\includegraphics[width=\textwidth]{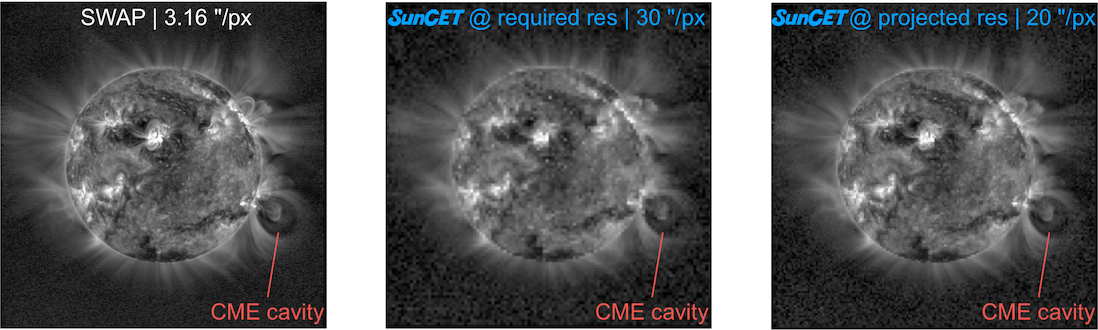}
\caption{CME cavity observed in PROBA2/SWAP 174~\AA\ binned down to SunCET required resolution of 30\arcsec\ (projected performance is 20\arcsec). The cavity remains easily identifiable. SunCET's SNR will be 9--30$\times$ higher off disk, making CME identification even easier. The 1.7~\Rs\ FOV shown here, the largest of any solar EUV imager to date, is SWAP's; SunCET's extends to 4 \Rs. Adapted from \citet{Byrne2014}.}
\label{fig:swap_cavity}
\end{figure}

SunCET requires spatial resolution better than 30\arcsec. CME flux ropes often manifest observationally as a cavity which trails behind a bright front \citep{Forsyth2006}. The smallest cavities have a diameter of 0.2 \Rs\ (180\arcsec) and are approximately circular, which corresponds to a circumference of $\sim$600\arcsec\ \citep{Fuller2009}. To account for non-circularities, we require $\sim$20 points outlining the cavity, which results in our spatial resolution requirement of 600\arcsec/20 = 30\arcsec. Figure \ref{fig:swap_cavity} shows a cavity observed in PROBA2/SWAP (3.16\arcsec\ resolution) binned down to demonstrate that cavities can be resolved at this resolution in practice.

\textit{Projected performance}: SunCET provides 20\arcsec\ resolution. Its plate scale is 4.8\arcsec/pixel so 2$\times$2 binning can be applied, which meets the Nyquist sampling criterion. 

\subsection{Mirrors}
\label{mirrors_filters}

\begin{figure}
\centering
\includegraphics[width=0.92\textwidth]{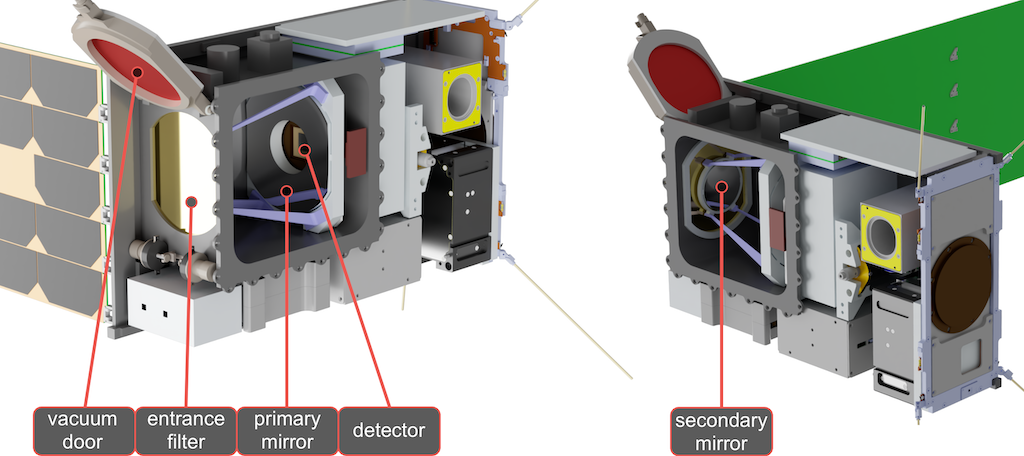}
\caption{SunCET's compact Ritchey-Chr\'etien telescope, which fits inside a 6U CubeSat with all typical bus components.}
\label{fig:cad_telescope}
\end{figure}

\begin{figure}
\centering
\includegraphics[width=0.8\textwidth]{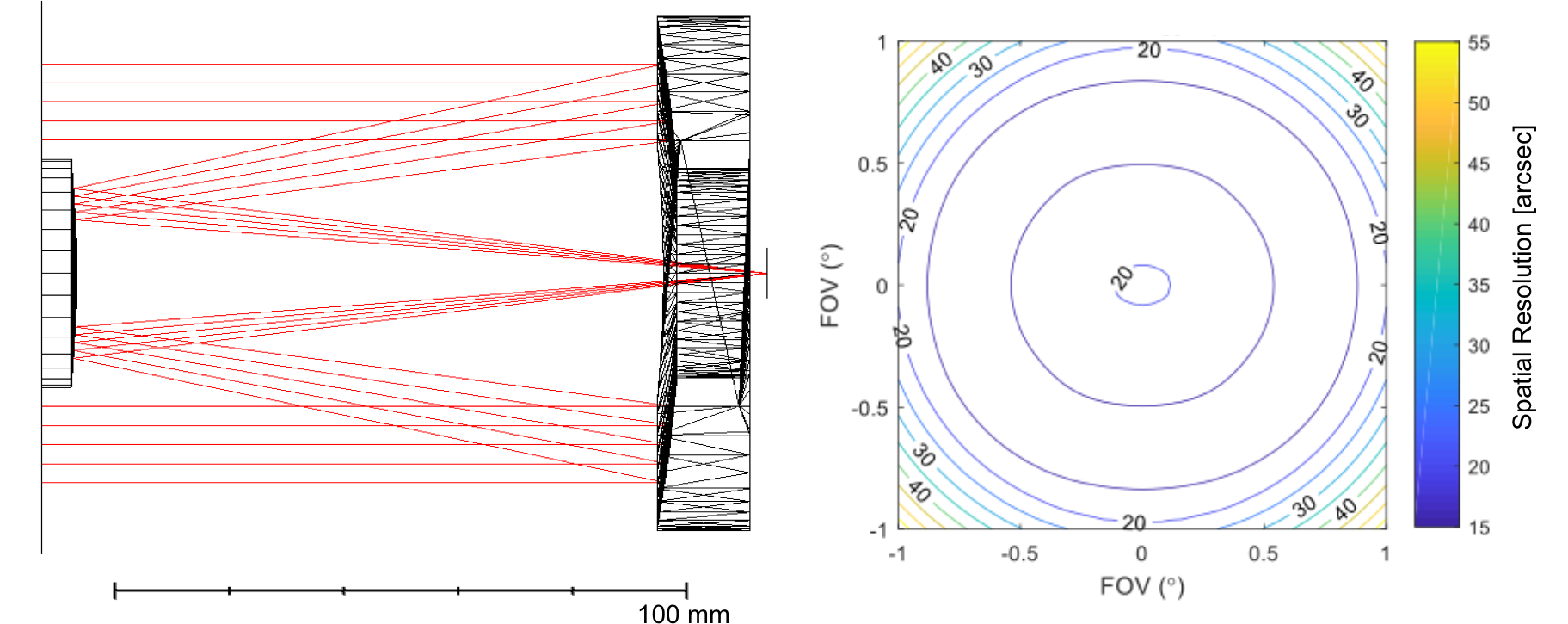}
\caption{Left: Ray trace of SunCET optics. Right: 80\% encircled spot diameter over the FOV. This simple design yields excellent performance, with a mean resolution of 20\arcsec\ that is flat across nearly the entire FOV.}
\label{fig:ray_trace}
\end{figure}

SunCET contains a Ritchey-Chr\'etien (RC) telescope encased in a vacuum chamber with a one-time-release door (Figure \ref{fig:cad_telescope}). This type of telescope has good performance for wide fields of view (Figure \ref{fig:ray_trace}) and has been used frequently for similar instruments (e.g., SOHO/EIT, STEREO/EUVI, GOES/SUVI). Despite its compact size, the telescope achieves nearly flat resolution across the wide FOV. The mount for the secondary mirror is designed with a coefficient of thermal expansion matching the mirror to account for focus sensitivity.

\subsection{Signal to Noise Ratio (SNR)}
\label{sec:snr}

SunCET requires a signal to noise ratio (SNR) $\geq$10. This is the international standard that defines digital image quality as ``acceptable" \citep{ISO12232}. The same standard defines SNR of 40 as ``excellent". These numbers are in line with the expectations of experts that have done CME image processing with coronagraph and EUV imager data. 

\begin{table}[htb]
\caption{SunCET SNRs for on-disk features and CME loops above the limb. Radiances are from GOES/SUVI 195~\AA\ images of the 2017-09-10 CME \citep{Seaton2018} and are extrapolated beyond its FOV of 1.7 \Rs. SNR at all heights is above the level that ISO 12232 defines as ``excellent". }
\label{tab:snr}
\resizebox{\textwidth}{!}{%
\begin{tabular}{|c|c|c|c|c|c|c|c|c|c|}
\hline
\rowcolor[HTML]{1A73C9} 
{\color[HTML]{FFFFFF} } & {\color[HTML]{FFFFFF} \begin{tabular}[c]{@{}c@{}}Quiet\\ Sun\end{tabular}} & {\color[HTML]{FFFFFF} \begin{tabular}[c]{@{}c@{}}Active\\ Region\end{tabular}} & {\color[HTML]{FFFFFF} Flare} & {\color[HTML]{FFFFFF} 1.05 \Rs} & {\color[HTML]{FFFFFF} 1.5~\Rs} & {\color[HTML]{FFFFFF} 2~\Rs} & {\color[HTML]{FFFFFF} 3~\Rs} & {\color[HTML]{FFFFFF} 3.5~\Rs} & {\color[HTML]{FFFFFF} 4 \Rs} \\ \hline
\begin{tabular}[c]{@{}c@{}}Radiance\\\relax [W/m$^2$/sr]\end{tabular} & 0.1 & 10 & 40 & 0.2 & \num{1.5e-2} & \num{3e-3} & \num{3e-4} & \num{1e-4} & \num{3e-5} \\ \hline
Effective exposure [s] & 0.025 & 0.025 & 0.025 & 0.025 & 10 & 10 & 10 & 10 & 10 \\ \hline
e$^-$/res-element & \num{1.48e4} & \num{1.48e6} & \num{5.94e6} & \num{2.97e4} & \num{8.9e5} & \num{1.78e5} & \num{1.78e4} & \num{5.94e3} & \num{1.78e3} \\ \hline
\begin{tabular}[c]{@{}c@{}}Saturation limit\\\relax [e$^-$/res-element]\end{tabular} & \num{1.08e5} & \num{1.08e5} & \num{1.08e5} & \num{1.08e5} & \num{1.08e6} & \num{1.08e6} & \num{1.08e6} & \num{1.08e6} & \num{1.08e6} \\ \hline
SNR & 122 & Saturated & Saturated & 172 & 944 & 422 & 133 & 77 & 42 \\ \hline
\end{tabular}%
}
\end{table}

\textit{Projected performance}: Table \ref{tab:snr} shows the SunCET SNR as a function of distance from the sun,  based on the parameters shown in Table \ref{tab:instrument_response}. Conservative radiance estimates come from GOES/SUVI 195 ~\AA\ images of a CME that was tracked all the way to the edge of the SUVI 1.7~\Rs\ FOV \citep{Seaton2018}. For the solar disk, the effective exposure is the median of three 0.025-second images; for 1.05--4~\Rs, it is the median of ten 1-second exposures. This removes energetic particle tracks and, for the long exposure, increases the full-well saturation limit of the detector by a factor of 10. These conservative estimates show that SunCET CME measurements would have an excellent SNR of 42 even out at 4~\Rs.

\medskip

\begin{table}[h!]
\caption{SunCET instrument parameters needed to calculate SNR.}
\label{tab:instrument_response}
\resizebox{\textwidth}{!}{%
\begin{tabular}{|c|c|l|}
\hline
\rowcolor[HTML]{1A73C9} 
{\color[HTML]{FFFFFF} \begin{tabular}[c]{@{}c@{}}Instrument \\ parameter\end{tabular}} & {\color[HTML]{FFFFFF} Value} & {\color[HTML]{FFFFFF} Description} \\ \hline
Wavelength & 170--200~\AA & Broadband response defined by mirror coating \\ \hline
\begin{tabular}[c]{@{}c@{}}Aperture\\ size\end{tabular} & 44.9~cm$^2$ & \begin{tabular}[c]{@{}l@{}}9.6~cm diameter truncated on two sides\\ to a height of 7.62~cm and a~4.8 cm \\ diameter secondary mirror obscuring its center\end{tabular} \\ \hline
\begin{tabular}[c]{@{}c@{}}Weighted factor\\ for broadband\end{tabular} & 6.88 & \begin{tabular}[c]{@{}l@{}}7 emissions in the bandpass weighted by their\\ quiet-Sun intensity to the 195~\AA\ emission line\\ (see Table \ref{tab:bandpass_lines})\end{tabular} \\ \hline
Pixel size & 7 \mymu m $\times$ 7 \mymu m & e2v CIS115 datasheet and confirmed in house \\ \hline
Pixel array & 1500 $\times$ 1500 & Full array is 1504 $\times$ 2000; $\sim$5 rows dedicated to dark \\ \hline
FOV & 4 \Rs & Design FOV (requirement is 3.5~\Rs) \\ \hline
Plate scale & 4.8\arcsec/pixel & \begin{tabular}[c]{@{}l@{}}From pixel size, number of pixels, and FOV; \\ Note that 2$\times$2 binning will be applied, \\ resulting in  9.6\arcsec/resolution-element\end{tabular} \\ \hline
\begin{tabular}[c]{@{}c@{}}Optics\\ throughput\end{tabular} & 0.06 & \begin{tabular}[c]{@{}l@{}}2 mirrors with B$_4$C/Mo/Al coatings (0.35 each),\\ entrance Al/C filter (0.6) with 5~lpi filter mesh (0.98), \\ Al secondary/pinhole filter (0.85)\end{tabular} \\ \hline
Quantum yield & 18.3 e$^-$/ph & Average over 170--200~\AA\ bandpass \\ \hline
Dark noise & $<$0.08 e$^-$/pixel/sec & At -10\degree C, from LASP lab tests \\ \hline
Readout noise & 5 e$^-$/pixel & From LASP lab tests \\ \hline
Fano noise & 1.3 e$^-$/pixel & Fano factor of 0.1 for Si \\ \hline
Max read rate & \begin{tabular}[c]{@{}c@{}}0.1 sec (full frame)\\ 0.025 (up to 500 rows)\end{tabular} & In SunCET, 500 rows corresponds to 0--1.33 \Rs \\ \hline
\end{tabular}
}
\end{table}

\begin{figure}
\centering
\includegraphics[width=0.5\textwidth]{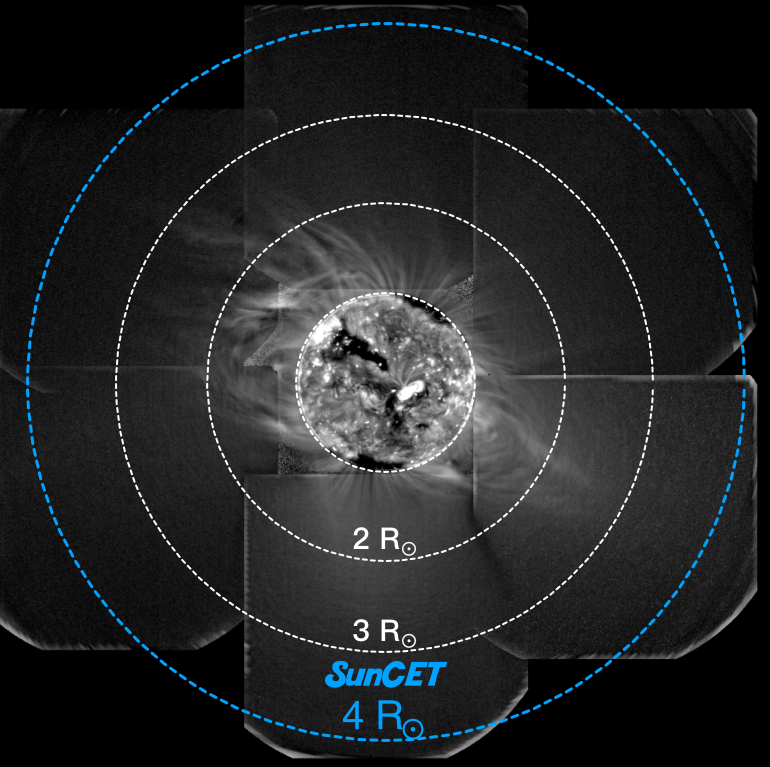}
\caption{Composite of GOES/SUVI 195~\AA\ off-point images that shows solar structure out to 3 \Rs\ --- even without a bright CME --- before straylight in the instrument becomes comparable with the coronal signal. Adapted from \citet{Tadikonda2019}.}
\label{fig:suvi_off_point}
\end{figure}

Few observations of the extended corona above $\sim$2~\Rs\ have been made in the EUV, but among these there is clear evidence that the CME signal will be detectable (\citealt{Tadikonda2019}, Figure \ref{fig:suvi_off_point}). At about 3~\Rs, noise in SUVI becomes comparable to solar signal. SunCET, however, is optimized for this large FOV. SunCET has a larger primary mirror geometric area (3.5$\times$), broadband wavelength response (6.88$\times$), and larger pixel solid angle (16$\times$) for a total 385$\times$ boost in signal compared to SUVI. Furthermore, the SunCET mirrors will be  polished to highest degree possible, up to 3 times the smoothness of SUVI's, to minimize scattered light. 

\section{Spectrograph Design}
\label{sec:spect_technical}
The SunCET irradiance spectrograph channel is a high-heritage off-Rowland circle design based on the SDO/EVE Multiple EUV Grating Spectrographs A2 (MEGS-A2) channel \citep{Crotser2007}. It provides the full-Sun solar irradiance from 170-340 \AA\ at 1 \AA\ spectral resolution. This EUV range is important for overlapping with the SunCET imager EUV bands for calibration purposes and provides additional science capability. It observes Fe IX through Fe XVI emission lines that often experience coronal dimming during CMEs \citep{Woods2011, Mason2014, Mason2016, Mason2019}. This allows for halo CME kinematics to be tracked even if SunCET is not deployed on multiple platforms with stereoscopic viewing angles. It also enables study of the energetics powering the CME as a function of time. It shares the vacuum door and detector with the SunCET imager, but has its own optical path including the entrance slit, filters, and grating. These measurements are especially pressing because EVE/MEGS-A experienced a CCD electronics anomaly in 2014 May, preventing the continued solar observations by MEGS-A. While other EVE channels and new GOES EUV Sensor (EUVS) channels are continuing solar EUV observations in the 170-340 \AA\ range, they are only broadband measurements that are not optimized for coronal dimming irradiance observations nor for detailed calibration of solar EUV imagers.

\subsection{Spectrograph Dynamic Range}
\label{sec:spect_dynamic_range}
The solar irradiance values, as measured from SDO/EVE \citep{Woods2012}, from 170-340 \AA\ range from \num{e-6}--\num{e-2} W/m$^2$/nm due to variations in the peaks of the emission line in this range, the reduced irradiance values between the strong emission lines, as well as solar activity including solar minimum times and during the largest solar flares; therefore, the required dynamic range of the spectrograph is \num{1e4}.

\textit{Projected performance}: The \num{8e6} dynamic range discussed in Section \ref{sec:dynamic_range} is more than two orders of magnitude better than needed for the spectrograph.

\subsection{Spectrograph Spectral Range and Resolution}
\label{sec:spect_spectral}

The SunCET spectrograph requires a spectral range between 170-340 \AA\ and 1 \AA\ spectral resolution.  The entrance to the spectrograph is a 3 $\times$ 0.028~mm in order to maximize the slit image height (cross-dispersion direction) on the allotted 500 pixel height of the detector to maximize the SNR, while the width is optimized to meet the 1 \AA\ spectral resolution requirement --- it is this slit width and the grating ruling that limits the spectral resolution. The grating ruling, distance and curvature are all optimized in order to meet the spectral range and resolution as well. 

The optical path after being dispersed from the grating goes through the hole in the secondary imager mirror and onto the common detector. The grating is a Type-I concave imaging grating in order to image the slit onto the detector. There is an Al/C entrance filter mounted to the entrance slit in order to limit the spectral bandpass close to the required range, and an additional Al filter prior for additional bandpass rejection at the entrance to the imager optical cavity as well as to reduce any stray light or pinholes that may develop in the first filter. 

Given the 1500 allotted pixels in the dispersion range, this gives a plate-scale resolution of approximately 0.11 \AA\ per pixel; therefore the spectrograph will oversample the spectral resolution by about a factor of 9$\times$, or 4.5$\times$ with the 2$\times$2 pixel binning. This allows for fits to spectral lines to be performed and allow for Doppler shift measurements of emission lines and plasma velocity flows during flares to be calculated \citep{Chamberlin2016, Hudson2011}

\textit{Projected performance}: SunCET provides 1 \AA\ spectral resolution across the fully observed 170-340 \AA\ spectral range. 

\subsection{Spectrograph Signal to Noise Ratio (SNR)}
\label{sec:spect_snr}

The SunCET spectrograph also requires a SNR of 10 or better as discussed in \ref{sec:snr}. This is achieved by using a long-slit and minimal optical elements, along with the high QE detector. The slit was also sized, and filter thickness optimized, to maximize the SNR without while conservatively not saturating or even go beyond the linear full well capacity of the the CMOS sensor. Even with a very large factor of 10 increase \citep{Chamberlin2008, Chamberlin2018} during flares for these lines given in Table \ref{tab:spect_snr}, they will still be almost another factor of 2 below the full-well of this sensor.

\begin{table}[htb]
\caption{The SunCET spectrograph SNRs for various strong emission lines. Irradiances are from SDO/EVE \citep{Woods2011}. SNR at all heights is above the level that ISO 12232 defines as ``excellent". }
\label{tab:spect_snr}
\resizebox{\textwidth}{!}{%
\begin{tabular}{|c|c|c|c|c|c|}
\hline
\rowcolor[HTML]{1A73C9} 
{\color[HTML]{FFFFFF} Wavelength (\AA)} & {\color[HTML]{FFFFFF} 171} & {\color[HTML]{FFFFFF} 193.5} & {\color[HTML]{FFFFFF} 195} & {\color[HTML]{FFFFFF} 304} & {\color[HTML]{FFFFFF} 335} \\ \hline
\begin{tabular}[c]{@{}c@{}}Irradiance\\\relax [W/m$^2$/sr]\end{tabular} & \num{6.7e-4} & \num{4.5e-4} & \num{6.3e-4} & \num{1.0e-3} & \num{1.0e-4} \\ \hline
Integration [s] & 10 & 10 & 10 & 10 & 10 \\ \hline
Counts/Pixel & 737 & 495 & 693 & 1100 & 110  \\ \hline
SNR & 272 & 237 & 282 & 444 & 145 \\ \hline
\end{tabular}%
}
\end{table}

\textit{Projected performance}: Table \ref{tab:spect_snr} shows the SunCET spectrograph SNR for five strong emission lines,  based on the parameters shown in Table \ref{tab:spect_instrument_response}. These  estimates show that SunCET solar spectral irradiance measurements would have an excellent SNR of better than 100.

\medskip

\begin{table}[h!]
\caption{SunCET spectrograph instrument parameters needed to calculate SNR.}
\label{tab:spect_instrument_response}
\resizebox{\textwidth}{!}{%
\begin{tabular}{|c|c|l|}
\hline
\rowcolor[HTML]{1A73C9} 
{\color[HTML]{FFFFFF} \begin{tabular}[c]{@{}c@{}}Instrument \\ parameter\end{tabular}} & {\color[HTML]{FFFFFF} Value} & {\color[HTML]{FFFFFF} Description} \\ \hline
Wavelength & 170--340~\AA & \begin{tabular}[c]{@{}l@{}}Contains various strong emission lines, \\including some that show coronal dimming.  \\Defined by grating equation. \end{tabular}\\ \hline
\begin{tabular}[c]{@{}c@{}}Aperture\\ size\end{tabular} & 0.0098 cm$^2$ & 3.0~mm tall $\times$ 28 \mymu m wide \\ \hline
\begin{tabular}[c]{@{}c@{}}Number of Pixels\\ per  emission line\end{tabular} & 2000 & \begin{tabular}[c]{@{}l@{}}500 pixels tall  $\times$ 4 pixels wide \\ (defined by slit) \end{tabular} \\ \hline
Pixel size & 7 \mymu m $\times$ 7 \mymu m & Teledyne e2v CIS115 datasheet and confirmed in house \\ \hline
Pixel allocation & 500 $\times$ 1500 & Full array is 1504 $\times$ 2000; $\sim$5 rows dedicated to dark \\ \hline
FOV & Full Sun & Solar Irradiance, image the slit \\ \hline
Plate scale & 0.011 nm & \begin{tabular}[c]{@{}l@{}}From pixel size, number of pixels, wavelength range; \\ Note: oversampling spectral resolution of 0.1nm \end{tabular} \\ \hline
\begin{tabular}[c]{@{}c@{}}Optics\\ throughput\end{tabular} & 0.0122 & \begin{tabular}[c]{@{}l@{}}Grating Efficiency (0.06), Pt Grating Coating (0.4), \\ Al/C entrance filter (0.6), Al secondary/pinhole filter (0.85)\end{tabular} \\ \hline
Quantum yield & 18.3 e$^-$/ph & Average over 170-200 \AA\ bandpass \\ \hline
Dark noise & $<$0.08 e$^-$/pixel/sec & At -10\degree C, from LASP lab tests \\ \hline
Readout noise & 5 e$^-$/pixel & From LASP lab tests \\ \hline
Fano noise & 1.3 e$^-$/pixel & Fano factor of 0.1 for Si \\ \hline
\end{tabular}
}
\end{table}

\section{Detector}
\label{sec:detector}

\begin{figure}
\centering
\includegraphics[width=0.30\textwidth]{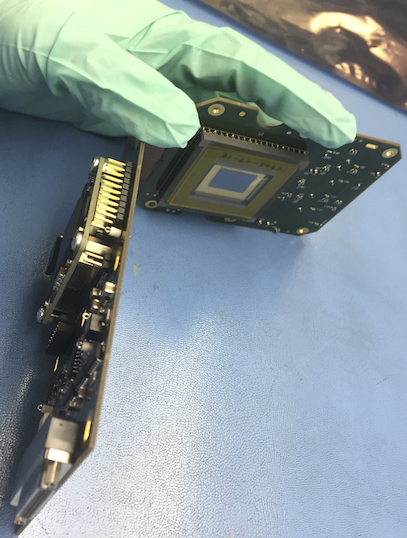}
\caption{The Teledyne e2v CIS115 detector and LASP Compact Camera and Processor (CCAP) that flew successfully on a NASA sounding rocket in 2018; CCAP is now flying on the CSIM CubeSat launched in 2018.}
\label{fig:detector}
\end{figure} 

SunCET uses a Teledyne e2v CIS115 back-illuminated, back-thinned CMOS sensor (Table \ref{tab:instrument_response}, Figure \ref{fig:detector}). This sensor is a 1504$\times$2000 pixel array, where a square area of 1500$\times$1500 pixels will be dedicated to the image while the remaining 500$\times$1500 pixels will record the spectrally dispersed slit image from the irradiance spectrograph. Using a single detector to record data from two technically different but scientifically complementary channels significantly reduces the technical resources needed while maximizing science potential.  

In 2017, LASP developed custom electronics for readout of this sensor that enables independent exposure control per row. A per-pixel readout is now being developed. LASP's ``Compact Camera and Processor" (CCAP; Figure \ref{fig:detector}) system with this detector was successfully flown in 2018 on the NASA 36.336 sounding rocket (PI: T. Woods, U. of Colorado/LASP) and more recently in January 2020 on the NASA 36.356 sounding rocket (PI: S. Bailey, Virginia Tech). CCAP includes a Xilinx Kintex-7 FPGA with an embedded 32-bit processor and dedicated image compression core.

\section{Instrument Requirements on Spacecraft}

The instruments described above place requirements on the performance and capabilities of whatever spacecraft hosts them. They are primarily driven by the imager. Pointing accuracy must be better than 30\arcsec with stability better than 30\arcsec\ RMS over 23 seconds and knowledge better than 10\arcsec. This ensures that the center of the sun stays in the center of portion of the detector dedicated to the imager and does not drift significantly during or between integrations. This pointing performance is achievable even on CubeSat platforms as demonstrated by the Miniature X-ray Solar Spectrometer (MinXSS), Arcsecond Space Telescope Enabling Research in Astrophysics (ASTERIA), Compact Spectral Irradiance Monitor (CSIM), and others \citep{Mason2017a, Pong2018}. Prime science data generation is heavily dependent on CME occurrence rates, but downlink schemes can easily be designed for flexibility and the ``poorest" CMEs can be ignored if there are bandwidth limitations. For CME occurrence rates at the middle of the rising phase of the solar cycle, SunCET generates $\sim$28 MB/day for the imager, and $\sim$65 MB/day of data for the spectrograph. These data are compressed using a lossless JPEG-LS scheme.

\section{Conclusions}

The SunCET instrument fills a crucial, historically under observed region of the Sun --- the middle corona --- precisely the region where CMEs experience the majority of their acceleration. This region is inherently very difficult to observe because of the extreme intensity dynamic range between the bright solar disk and the dim corona. SunCET introduces a new technology that avoids the limitations of previous instruments. By developing a detector that can vary exposure time across its surface, we can simultaneously observe the disk without saturating and the dim middle corona; allowing us to track CMEs from their initiation all the way through their primary acceleration phase. Moreover, we can image spectra on the same detector with their own, independent integration time. 

\begin{figure}
\centering
\includegraphics[width=0.6\textwidth]{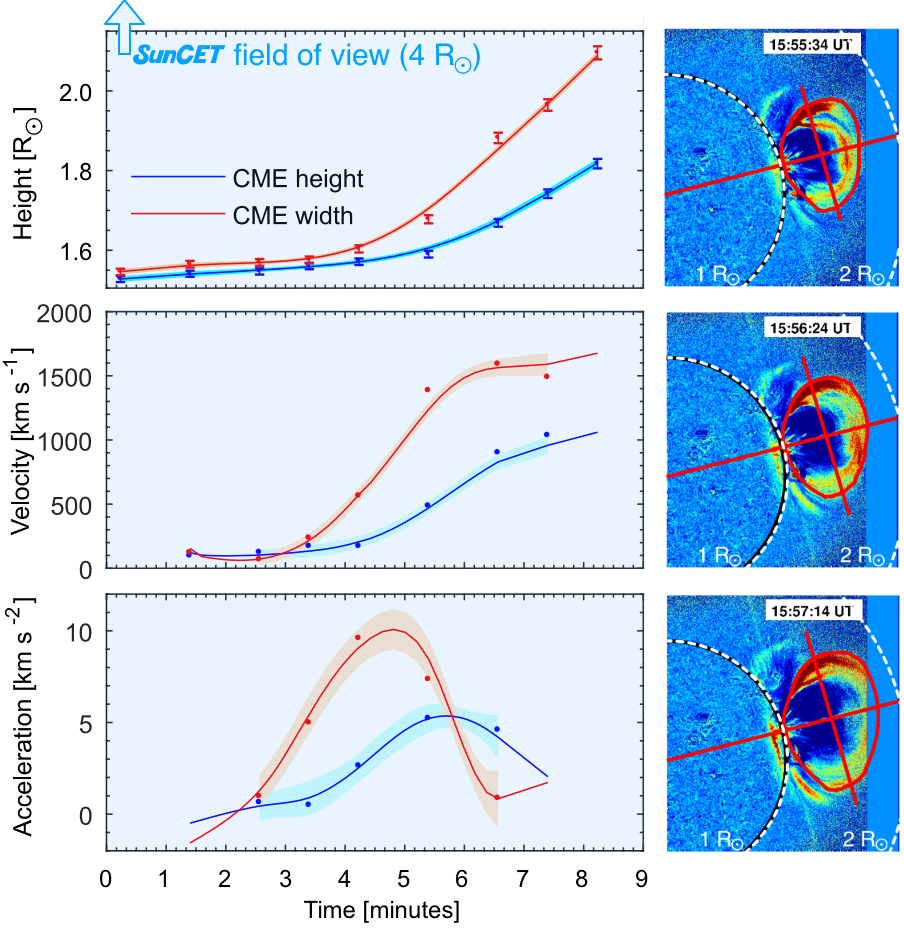}
\caption{Tracking a very fast CME in GOES/SUVI 195~\AA\ base-difference images. The CME quickly extended beyond the FOV of SUVI. SunCET's FOV (light blue shading) is more than twice as large. Adapted from \citet{Veronig2018}.}
\label{fig:suvi_height_time}
\end{figure}

There is a large body of knowledge for tracking CMEs in coronagraphs and EUV imagers \citep{Sarkar2019, OHara2019, Veronig2018, Byrne2014, Mierla2013, Bein2011, Gopalswamy2009, Vrsnak2007}. SunCET data processing will employ the techniques already developed for other observatories but improve the results because of its wider FOV (e.g., \citealt{Veronig2018}; Figure \ref{fig:suvi_height_time}) and that it does not require the serendipitous alignment between instrument off-point campaigns and CME occurrence (e.g., \citealt{OHara2019}).

Below we summarize:

\begin{enumerate}
  \item The majority of CME acceleration occurs in a historical observational gap: the middle corona
  \item Observations of full CME acceleration profiles provide tight constraints on models and thus our physical understanding of how the magnetically-dominated corona influences CME kinematics
  \item SunCET provides these observations, overcoming the limits of traditional technologies with a novel simultaneous high-dynamic-range detector
  \item SunCET is compact and thus suitable for CubeSat missions or an instrument on a larger spacecraft
\end{enumerate}

SunCET is presently in a NASA-funded, competitive Phase A as a 6U CubeSat and has also been proposed to NASA as an instrument onboard a 184 kg Mission of Opportunity.

\section{Acknowledgements}
      J.P.M thanks the numerous people who contributed to the development of the SunCET concept design and the reviewers for their commments that made this paper stronger. A.M.V. and K.D. acknowledge the Austrian Space Applications Programme of the Austrian Research Promotion Agency FFG (ASAP-11 4900217 CORDIM and ASAP-14 865972 SSCME, BMVIT).

\bibliography{main}

\end{document}